\documentclass[LaTeX,twocolumn,epjc3]{svjour3}          
\usepackage[linktocpage,breaklinks]{hyperref}

\usepackage{float}  
\usepackage{amsmath}
\usepackage{cuted}
\usepackage{xcolor}  
\definecolor{romared}{RGB}{142,0,28}
\hypersetup{colorlinks=true,
            citecolor=romared,
            linkcolor=romared,
            urlcolor=romared}
\RequirePackage[T1]{fontenc}

\smartqed  
\RequirePackage{graphicx}
\RequirePackage{mathptmx}      
\RequirePackage{flushend}
\RequirePackage[numbers,sort&compress]{natbib}

\journalname{Eur. Phys. J. C}

\begin{document}

\title{Time evolution of perturbations in quasi-Schwarzschild black holes}

\author{Orival R. de Medeiros\thanksref{e1,addr1}
        \and
        Mateus Malato Corrêa\thanksref{e2,addr1}
        \and Caio F. B. Macedo\thanksref{e3,addr1,addr2}
}

\thankstext{e1}{orival@ufpa.br}
\thankstext{e2}{malato.mateus@gmail.com}
\thankstext{e3}{caiomacedo@ufpa.br}

\institute{Programa de Pós-Graduação em Física, Universidade Federal do Pará, 66075-110, Belém, PA, Brazil\label{addr1}
          \and
          Faculdade de Física, Campus Salinópolis, Universidade Federal do Pará, 68721-000, Salinópolis, Pará, Brazil\label{addr2}
          }

\date{Received: date / Accepted: date}

\maketitle

\begin{abstract}
Parametric deviations of quasinormal modes~(QNMs) is a common feature of beyond General Relativity (GR) theories. For theories with additional degrees of freedom, such as scalars and vectors, new family of modes might appear, usually called scalar-led and vector-led modes. Although a power series expansion in terms of the new parameters entering the potential is usually suitable to describe the frequency of the modes, the time-evolution of signals might present a richer structure, with different behavior in the tail, the presence of new modes (such as massive modes), or even instabilities. All these features are not explicitly exposed by a pure frequency domain analysis and might give hints of new physics. In this paper, we investigate the time evolution of signals considering potentials that slightly deviate from the ones coming from GR, looking into scalar, vector and metric perturbations. We focus on deviations that can be parametrized by a sum of $\sim 1/r^j$ terms in the effective potential, analyzing the effect of each term on the time domain profile.  
\end{abstract}

\section{\label{sec:introduction}Introduction}

Black holes (BHs), some of the most intriguing and extreme entities in the cosmos, arise as a direct prediction of GR, which provides a robust mathematical foundation for understanding the nature of spacetime (see Ref.~\cite{chandrasekhar1998mathematical} for a robust account on BH physics).  BH solutions not only deepen our understanding of spacetime but also serve as natural laboratories for testing GR and exploring possible deviations from it~\cite{Berti:2015itd,Stairs:2003eg}. 

Einstein's GR predicted the existence of gravitational waves, which can be generated by various astrophysical events, including the merger of two BHs. We now have around a hundred detections from binary systems \cite{KAGRA:2021vkt,abbott2016observation,abbott2016gw151226,abbott2017gw170814,abbott2017gw170104,abbott2017gw170817}, with particularly loud signals coming from the merger of BHs. The coalescence of two BHs can be divided into three distinct phases: (i) inspiral, where the BHs lose energy through gravitational radiation and gradually approach each other; (ii) merger, where the event horizons coalesce into a single BH; and (iii) ringdown, during which the remnant BH settles into a stationary configuration by emitting gravitational waves. This final stage is analogous to the behavior of a perturbed physical system, in which the dominant oscillations governing energy dissipation are known as QNMs. At very late times, the signal transitions from a QNM-dominated phase to a regime where power-law or exponential tails govern the decay, a behavior that can be understood in the context of Price’s Law.~\cite{PricePhysRevD.5.2419,Price:1972pw,Cunningham:1978zfa,Gomez:1992fk,Gundlach:1993tp,Gundlach:1993tn,Leaver:1986gd,Ching:1994bd,Andersson:1996cm}.

GR has successfully explained gravitational phenomena in the weak-field regime. The theory is compatible with all experimental tests such as the precession of Mercury's perihelion, the deflection of starlight, and time dilation~\cite{will2014confrontation,hafele1972around,Bertotti:2003rm}. However, it is possible that GR needs to be modified in the strong-field gravity regime~\cite{silva2024quasinormal,Endlich:2017tqa,Yunes:2013dva,Ishak:2018his}. It is known that GR is plagued by the existence of singularities and the modifications of it may prevent them~\cite{Parker:1973qd,Parker:1990mk}. Cosmological measurements also raise important conceptual questions, such as why the value of the cosmological constant is so small and why its energy density is so close to the current matter density, which could be explained in alternative views of gravity. Within GR, it is not possible to find a dynamical solution to the cosmological constant problem~\cite{Clifton:2011jh,Berti:2015itd}.

In the coalescence of two BHs, the ringdown consists of the relaxation of the newly formed merger remnant toward its equilibrium state. The ringdown approximately corresponds to the time interval during which the gravitational waveform can be described as a sum of exponentially damped with unique frequencies and damping times, known as QNMs~\cite{chandrasekhar1975quasi,berti2009quasinormal}. The seminal study that uncovered the ringdown characteristics was conducted by Vishveshwara, by investigating the time evolution of gravitational pulses in Schwarzschild spacetime~\cite{vishveshwara1970scattering}, a methodology that is still used up this day. QNMs can serve as a test for GR since the no-hair theorem implies that (vacuum) stationary BH spacetimes are characterized by their mass and angular momentum, being uniquely described by the Kerr spacetime~\cite{berti2006gravitational,meidam2014testing,israel1967event,israel1968event,Berti:2018vdi}. 
Therefore, one way to test possible modifications is to analyze the behavior of perturbations of BHs during the ringdown phase, where the gravitational response is dominated by QNMs. 

Beyond the ringdown, alternative theories of gravity can also affect the late-time behavior of signals, which is not directly captured by the spectrum alone. In this paper, we study the time evolution of initial profiles in spacetimes that slightly deviate from the Schwarzschild BH, following the approach used in Refs.~ \cite{cardoso2019parametrized,Hirano:2024fgp}. We analyze scalar, vector, and gravitational field perturbations. We demonstrate that, in addition to parametric deviations in the ringdown phase, late-time modifications arise depending on how the potential is altered, i.e., on the specific alternative theory being considered. Notably, two key differences emerge: the presence of additional oscillatory patterns induced by an effective mass term (similarly to~\cite{Burko:2004jn}) and an oscillatory behavior in the tail. Furthermore, when the modifications to the potential are negative, instabilities may be triggered.

The paper is structured as follows. In Sec.~\ref{sec:Framework}, we introduce the master equations that describe the perturbations and discuss the structure of the modified potential. In Sec.~\ref{sec:timeevolution}, we analyze a frequency domain computation, highlighting the impact of the modified potential. We also outline the time evolution method used in our results. Sec.~\ref{sec-4-cap-2}
 presents our main results on the time evolution of perturbations and the excitation of QNMs. Finally, we summarize our conclusions and discuss potential extensions of this work in Sec.~\ref{sec:conclusion}. Throughout this paper, we use natural units $G=c=1$.


\section{\label{sec:Framework}Framework}

In a Schwarzschild BH spacetime background, the equation governing the vibration modes for any induced disturbance, being scalar ($s=0$), vector ($s=1$), or gravitational fields ($s=2$), is known as the master equation, being
\begin{equation}
    \label{eq:MasterEquation}
    \frac{\partial^2\Psi^2_{\ell}}{\partial r^2_*}-\frac{\partial^2\Psi^s_{\ell}}{\partial t^2} - V_0\Psi^s_{\ell} = 0,
\end{equation}
where $V_0$ is the effective potential, being~\cite{berti2009quasinormal}\footnote{In the context of gravitational perturbations, the modes of oscillation are described by two types of equations: polar (even-parity, or Zerilli) and axial (or odd-parity, or Regge-Wheeler). These two are related through a transformation and are isospectral. Here we focus only on the Regge-Wheeler potential.}
\begin{equation}
    \label{eq:PotentialEffective_s01}
    V_0 = f\left(\frac{\ell(\ell+1)}{r^2}+\frac{2M(1-s^2)}{r^3}\right),
\end{equation}
and $r_*$ is the tortoise coordinate, defined by $dr_*=dr/f(r)$.

Given a modified theory of gravity that slightly deviates from GR, we introduce its parametrization in the master equations by incorporating power-law corrections into the effective potential:
\begin{eqnarray}
    \label{eq:PotencialModificado}
    V &=& V_0 + \delta V,~{\rm with}\\
    \label{eq:DeltaVs}
    \delta V &=& \frac{f}{(2M)^2} \sum_{j=0}^{\infty}\beta_j\left(\frac{2M}{r}\right)^j,
\end{eqnarray}
where $\beta_j$ are constant coefficients. While the above procedure seems to be artificial, there are some alternative theories of gravity or even more general solutions within GR that fits the potential~\eqref{eq:PotencialModificado}. We refer the readers to Ref.~\cite{cardoso2019parametrized} where some examples are provided.

In the following, we investigate Eq.~\eqref{eq:MasterEquation} with the effective potential being described by \eqref{eq:PotencialModificado}. We carry some of the calculations generically, and in the numerical results, we shall look into only one non-vanishing $\beta_j$. In practice, this means that in some cases
\begin{eqnarray}\label{eq:potencial_simpli}
    V=\left(1-\frac{2M}{r}\right)\left[\frac{\ell(\ell+1)}{r^2}\right.&+&\left.\frac{2M}{r^3}(1-s^2)\right.\nonumber\\
    &+&\left.\frac{\beta}{(2M)^2}\left(\frac{2M}{r}\right)^j\right],
\end{eqnarray}
specifying the value of an integer $j$. Notice that in this case we dropped the $j$-index on $\beta$ as it is redundant. This allows us to characterize the effect of each individual term on the potential expansion.

\begin{figure}[h!]
    \centering
        \includegraphics[width=1\linewidth]{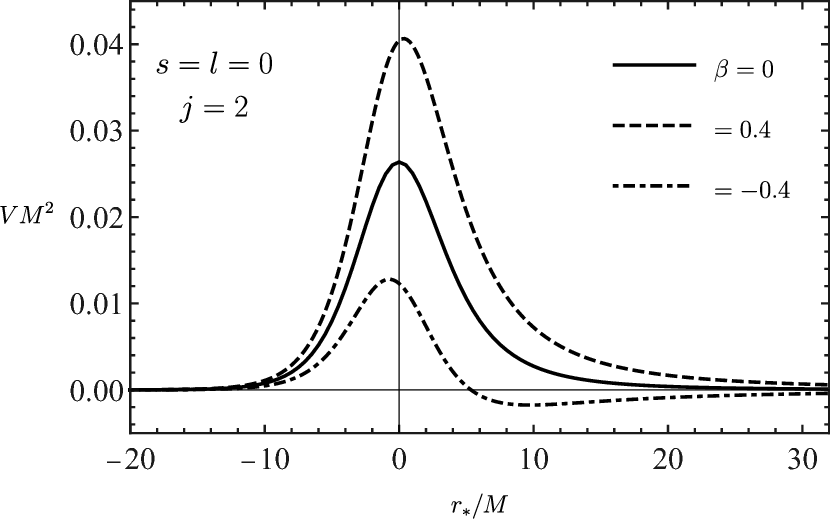}
        \caption{Effective potential given by Eq.~(\ref{eq:potencial_simpli}), for $j=2$ and different values of $\beta$. Cases with $\beta>0$ generate a potential barrier greater than the Schwarzschild case ($\beta=0$), while for $\beta<0$, may allow a region where the potential is negative.}
        \label{FIG:Pot_j2_ls0_diffbeta}
\end{figure}

\begin{figure}[h!]
    \centering
        \includegraphics[width=1\linewidth]{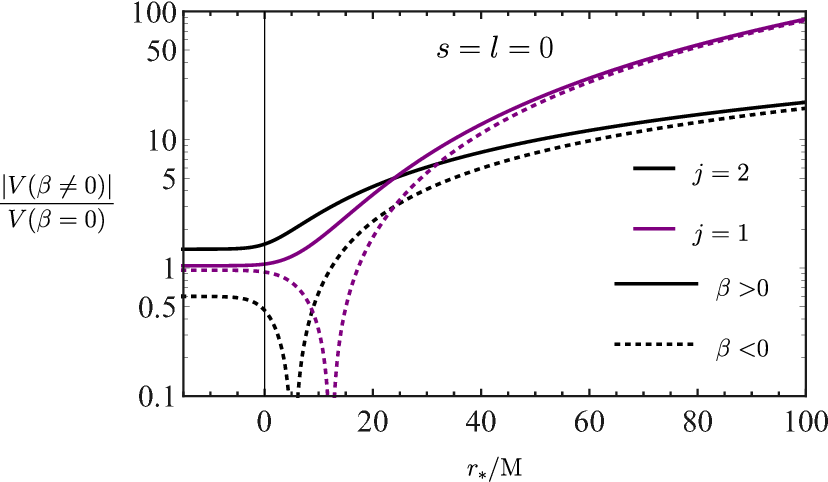}
        \caption{The potential normalized by the Schwarzschild case for $s=l=0$. In the case $j=1$, we have $\beta =\pm0.004$, and in $j=2$, we have $\beta=\pm0.4$. The slope for increasing $r_{*}$ is related to the tail in the time evolution, and for $\beta<0$ we have $V(\beta<0)<0$ for some values of $r_{*}$. }
        \label{fig:RatioPot_j1e2_ls0_diffbeta}
\end{figure}

In Fig.~\ref{FIG:Pot_j2_ls0_diffbeta}, we plot the potential for $s=\ell=0$, $j=2$, and $\beta=0, \pm0.4$, which are exaggerated values for illustration purposes. Increasing the value of $\beta$ leads to a potential that is $V(\beta>0)>V(\beta=0)$, in all spacetime, meaning a higher potential barrier that is related to the real part of the QNM. In the case of negative $\beta$, we can have a region where the potential is negative, which is related to possible instabilities, as we show in Sect.~\ref{subsec:instability}. Different values for $s,\ell$, present similar results to those outlined here for $s=l=0$, the main change is the negative part of the potential, which needs higher values for $|\beta|$.

In Fig.~\ref{fig:RatioPot_j1e2_ls0_diffbeta}, we plot the module of potential normalized by the Schwarzschild case. The purple lines are for $j=1$ and $\beta=\pm 0.04$, while in the black lines, we have $j=2$ and $\beta=\pm 0.4$. The dotted lines are for $\beta<0$. The solid lines are for $\beta>0$, and their behavior for $r_{*}\gg 0$, is related to the late time tails since this is determined due to backscattering by the potential far from its maximum, and thus the presence or not of oscillations in the tail is determined by this region. In the results, we show the evolutions considering the values depicted here and we see a different behavior than the GR case on the late-time tail.

\section{\label{sec:timeevolution}Frequency domain and time domain}

\subsection{Frequency domain calculations and the spectra}

Before diving into the time-domain, it is worth to introduce the QNM frequency computations. Here we employ the Leaver method, also known as continued fraction (CF), due to its accuracy~\cite{Leaver:1985ax,Nollert:1993zz}. Some of these computations were already presented in Ref.~\cite{cardoso2019parametrized}, but using a direct integration method. We focus on the case in which the potential corrections starts at $j>0$. The inclusion of the $j=0$ term modifies the dispersion relation of the fields, which causes a change in its boundary conditions.

Consider a time decomposition of the form $\Psi^s_\ell(t,r)=\psi^s_\ell(r)e^{-i\omega t}$. We obtain from \eqref{eq:MasterEquation}
\begin{equation}
 \frac{d^2\psi^s_\ell}{dr_*^2}+(\omega^2-V)\psi^s_\ell=0.
 \label{eq:frequencyd}
\end{equation}
As the potential goes to zero at the boundaries for $j>0$ (horizon and infinity), the physical conditions for QNMs are given by
\begin{equation}
 \psi^s_\ell(r\to 2M)\approx e^{-i\omega r_*},~{\rm and}~\psi^s_\ell(r\to \infty)\approx e^{i\omega r_*}.
\label{eq:boundary}
\end{equation}
The above boundary conditions define an eigenvalue problem for $\omega$, which corresponds to a discrete set of QNM frequencies.

We have that a solution obeying the required boundary conditions~\eqref{eq:boundary} can be written in terms of a series of the form
\begin{equation}
    \psi^{s}_\ell=\sum_{n=0}^{\infty}a_n e^{i \omega r} \left(\frac{r-2 M}{r}\right)^n \left(\frac{r}{2 M}-1\right)^{-2 i M \omega } \left(\frac{r}{2 M}\right)^{4 i M \omega }.
\end{equation}
By replacing the above series into the differential Eq. \eqref{eq:frequencyd}, we can obtain generically a $j_{\rm max}+1$-term recurrence relation, where $j_{\rm max}$ is the upper limit on the summation of the effective potential\footnote{For $j_{\rm max}<4$ one can construct a recurrence relations with less terms, as the expansion of the potential matches the power in $r$ with the ones coming from GR.}. For instance, for $j_{\rm max}=4$, we get the following 5-term recurrence relation for the coefficients

\begin{align}
A_0 a_{1}+B_0 a_0=0,\\
A_1 a_{2}+B_1 a_1+C_1 a_{0}=0,\\
A_2 a_{3}+B_2 a_2+C_2 a_{1}+D_2 a_{0}=0,\\
A_n a_{n+1}+B_n a_n+C_n a_{n-1}+D_n a_{n-2}+E_n a_{n-3}=0,
\end{align}

where the last equation is valid for $n\geq3$ and
\begin{align}
    A_n &= -(n + 1)(-4iM\omega + n + 1), \\
    B_n &= \beta_1 + \beta_2 + \beta_3 + \beta_4 - 32M^2\omega^2 \notag \\
        &\quad - 4iM(5n + 2)\omega + 3n^2 + 2n - s^2 + \ell^2, \\
    C_n &= -\beta_2 - 2\beta_3 - 3\beta_4 + n(2 + 24iM\omega) \notag \\
        &\quad + 8M\omega(6M\omega - i) - 3n^2 + 2s^2 - \ell(\ell + 1), \\
    D_n &= \beta_3 + 3\beta_4 + (-4iM\omega + n - s - 1) \notag \\
        &\quad \times (-4iM\omega + n + s - 1), \\
    E_n &= -\beta_4.
\end{align}

The above recurrence relation can then be reduced to a three-term through $j_{\rm max}-2(=2$, in the case of $j_{\rm max}=4$) Gaussian elimination steps, which can be recursively solved with the standard methods~\cite{Pani:2013pma}, finding the frequencies $\omega=\omega_r+i \omega_i$. Other cases with $j_{\rm max}>4$ follows in a similar fashion.

\begin{figure*}
    \centering
    \includegraphics[width=0.3\linewidth]{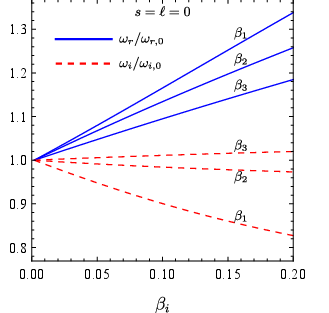}\includegraphics[width=0.3\linewidth]{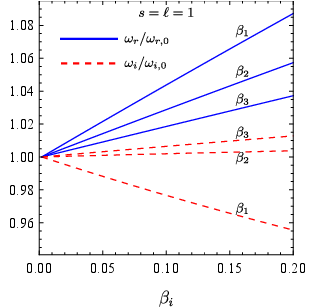}\includegraphics[width=0.3\linewidth]{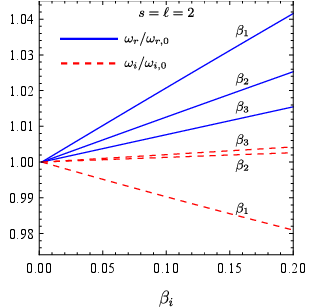}
    \caption{Real and imaginary parts of the QNM frequencies, normalized by the values in the Schwarzschild spacetime. We show the corrections with $\beta_{i}>0$ due to $j=1,\,2$ and $3$, for $s=\ell=0$ (left panel), $s=\ell=1$ (middle panel), and $s=\ell=2$ (right panel).}
    \label{fig:modes}
\end{figure*}

Focusing on a single $\beta_j$, the results in the frequency domain is shown in Fig.~\ref{fig:modes}. It is clear that, in agreement with Ref.~\cite{cardoso2019parametrized}, the QNM frequency deviates parametrically for small values of $\beta$, with a greater deviation occurring for $s=\ell=0$ case. A similar result holds for the overtones, as can be seen from Ref.~\cite{Hirano:2024fgp}. 

It is known that additional terms in the potential might cause instabilities. For example, when one considers certain alternative theories of gravity with high-curvature corrections coupled to additional fields, vacuum black holes can spontaneously create hair~\cite{Silva:2017uqg,Doneva:2017bvd,Antoniou:2017acq}. In our case, found no unstable modes for $\beta_j>0$. However, for negative enough values of $\beta_j$ instabilities appear. We were able to find these modes with the Leaver method presented here, but our computation loses accuracy as the mode crosses ${\rm Re}(\omega)$ axis. Nonetheless, in the time-evolution we illustrate clear evidence for the unstable modes, with a direct comparison with a frequency domain calculation. 

\subsection{\label{sec:level2}Time evolution of fields and QNMs excitation}

To find the time-domain response of an initial pulse, we assume that at $t = t_0$, the initial condition is a static Gaussian centered at $r_* = x_0$, that is
\begin{equation}
	\label{eq:GaussianPackage}
	\Psi(r_*,t_{0}) = ae^{-b(r_* - x_0)^2}
\end{equation}
and
\begin{equation}
	\label{eq:InitialConditionGaussianPackage}
	\frac{\partial\Psi(r_*,t)}{\partial t}\Bigg\vert_{t=t_0} = 0.
\end{equation}
Here we consider $a = 0.15$, $b = 0.05$, $x_0 = 50$, and the extraction point (observer) is located at $r_* = 50$. The ringdown response is not sensitive to the initial conditions. 

We solve the wave equation through a method of lines, uniformly discretizing the radial tortoise coordinate $r_*$, and solving  the ordinary differential equations in the time coordinate each point of the grid through a Mathematica NDSolve code, using the initial conditions~\eqref{eq:GaussianPackage} and~\eqref{eq:InitialConditionGaussianPackage}. As usual in the time evolution of wave functions, we extract the solution at some external point and track the time dependence of the pulse there. In order to avoid any contamination from the boundaries (horizon and infinity), the values for the minimum and maximum $r_*$ on the integration grid are pushed far from the extraction point, such that they are causally disconnected from the extracted signal. We have also performed a consistency check by decreasing the grid spacing, verifying convergence of the solutions.

To illustrate the evolution of the fields in Fig.~\ref{Figura:Ringdown3modos} we show the GR case, where we represent the dominant multipole for each type of perturbation, with $s = \ell= 0$ (left panel), $s =\ell = 1$ (middle panel), and $s = \ell = 2$ (right panel). We characterize each signal by three specific regions: $(i)$ an initial prompt that depends on the initial conditions; $(ii)$ the ringdown phase, in which the field is in a damped sinusoid state determined by the QNMs; and $(iii)$ in this final stage, the field reaches a more stable state and the oscillation decreases significantly, a state of relative equilibrium represented by the tail
that is an advanced stage after collapse~\cite{gundlach1994late,gundlach1994lateII}.

\section{Evolution of fields in quasi-Schwarzschild}
\label{sec-4-cap-2}
\hspace{0.5cm}

As mentioned before, the form of the quasi-Schwarzschild effective potential, given by Eqs.~\eqref{eq:PotencialModificado}--\eqref{eq:DeltaVs}, suggests that it is instructive to separate the dependency of each term in the power series individually. An effect established in Ref.~\cite{cardoso2019parametrized} and reviewed in the previous section is the parametric deviation of the QNM, which gives a dependence of the modes in relation to the extra parameters that can be expanded to a power series. We should expect, therefore, that in all cases we will have changes in the ringdown stage of the time evolution. However, the analysis in~\cite{cardoso2019parametrized} does not capture the existence of possible additional modes or changes in the tail.

To reinforce, the evolutions presented in this section have the following initial data parameters $(x_0/M,a,M^2b) = (50,0.15,0.05)$, where the wave function is extracted at $r_*=50M$. To better understand how the perturbations evolve, let us now turn to the expression of the effective potential given by Eq.~(\ref{eq:potencial_simpli}), which corresponds to Eq.~(\ref{eq:PotencialModificado}) without the summation.

In what follows, we will consider the evolution of perturbations subject to the effective potential (\ref{eq:potencial_simpli}) using different values of $s$ and $j$. For the values of $\ell$, we consider only the cases where $\ell = s$, which usually dominate the signals emitted by physical sources.
For each case, we will choose different intervals of $\beta$ with the aim of having corrections large enough for the effects of the terms to be highlighted without possible interference from effects that appear on different scales.

\subsection{Terms $j=0$ and the presence of the spectrum quasi-bound}

\begin{figure*}
    \centering

        \centering
        \includegraphics[width=0.33\textwidth]{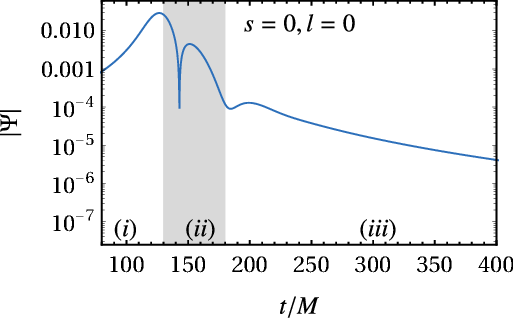}
        \includegraphics[width=0.33\textwidth]{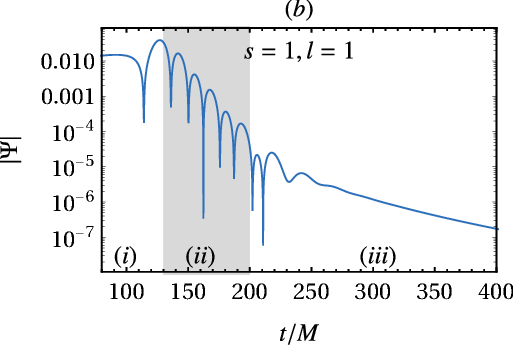}
        \includegraphics[width=0.33\textwidth]{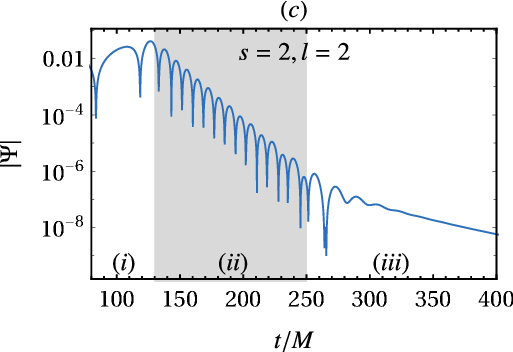}
        \caption{For gaussian packets, selected the following constants: $a = 0.15$, $b = 0.05/M^2$ starting at $t_0 = 0$ s in $x_0 = 50M$. With the observer at $r_* =50 M$. The three cases above correspond to $s = 0, 1$ and $2$ and vanishing $\beta$. The shaded region indicates the ringdown stage. }
  \label{Figura:Ringdown3modos}

    \vspace{1em}

        \centering
        \includegraphics[width=0.33\textwidth]{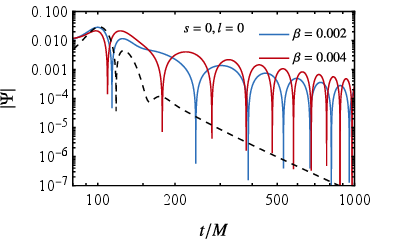}
        \includegraphics[width=0.33\textwidth]{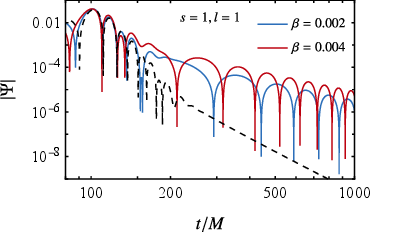}
        \includegraphics[width=0.33\textwidth]{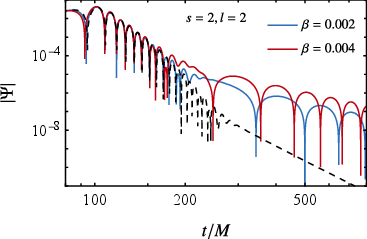}\\
        \includegraphics[width=0.33\textwidth]{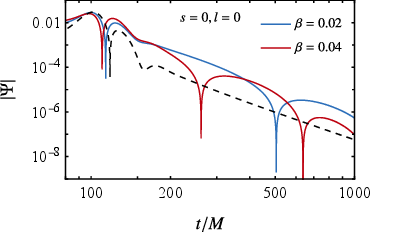}
        \includegraphics[width=0.33\textwidth]{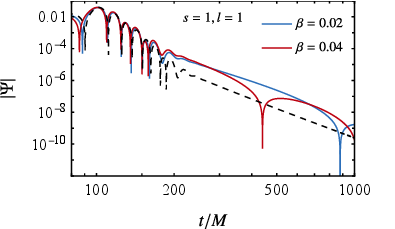}
        \includegraphics[width=0.33\textwidth]{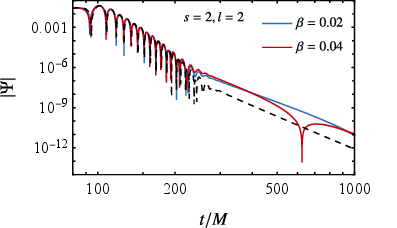}\\
        \includegraphics[width=0.33\textwidth]{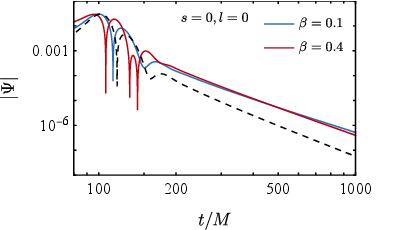}
        \includegraphics[width=0.33\textwidth]{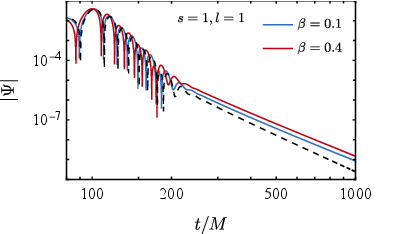}
        \includegraphics[width=0.33\textwidth]{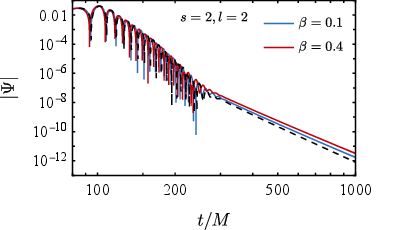}\\
        \includegraphics[width=0.33\textwidth]{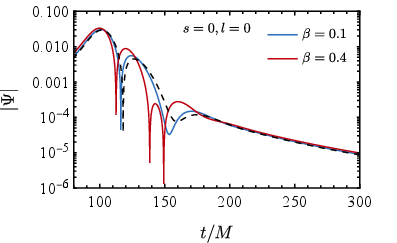}
        \includegraphics[width=0.33\textwidth]{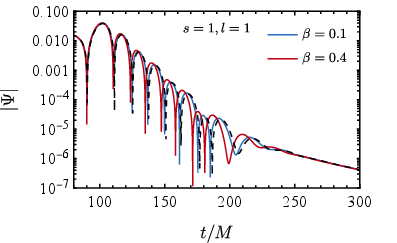}
        \includegraphics[width=0.33\textwidth]{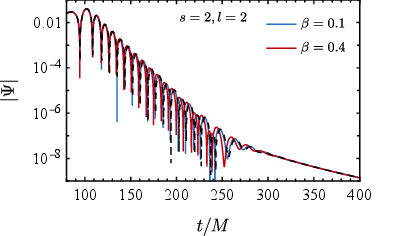}
        \caption{TTemporal evolution of a Gaussian perturbation. Each panel illustrates the temporal evolution of the scalar, vector, and gravitational fields, represented by  $s = \ell =  (0, 1, 2)$, respectively. The rows correspond to the values of $j$, ranging from 0 to 3, organizing the evolution of the fields in a clear and systematic structure. Dashed line denotes the result of Schwarzschild evolution with $\beta = 0$ represented in Fig.~\ref{Figura:Ringdown3modos}.}
        \label{fig:j_evol}

\end{figure*}


Let us first consider $j=0$. From the Eq.~\eqref{eq:potencial_simpli}, we see that, for this value of $j$, the deviation in the effective potential is essentially a mass term given by $\mu^2=\beta/(2M)^2$. Therefore, we can resort on the literature on massive fields to explain the effects of this additional term ~\cite{Koyama:2001ee,Koyama:2001qw,Burko:2004jn}.

For massive fields, there are two distinct spectra, essentially related to the asymptotic behavior of perturbations, called photon sphere modes (PSMs) and quasi-bound modes (QBMs).

PSMs exist for the case $\beta = 0$, i.e. in Schwarzschild, and are related to null trajectories, and an approximate expression can be obtained even by geodesic parameters~\cite{Cardoso:2008bp}, hence the name. These are the modes that appear in the ringdowns observed in Fig.~\ref{Figura:Ringdown3modos} and are related to the parametric expansion considered in the Ref.~\cite{cardoso2019parametrized} and, hence, this characteristic will be partially shared by the ringdowns of the evolutions considered here.

QBMs have the characteristic of trapping perturbations due to the exponential decay for large $r$, as a consequence of the mass term $\mu^2\sim\beta/(2M)^2$. An approximate expression of the modes can be seen in Refs.~\cite{Detweiler:1980uk} and is analogous to the bound states of the hydrogen atom, where the imaginary part exists due to the dissipative effect of the existence of the horizon.

The main effect of the mass term and the QBMs on the time evolution is the change in the signal at late times, assuming small values of $\beta$. Without the decay of perturbations, the signal stays trapped for a longer time, oscillating and decaying slowly, with a frequency that scales with the mass of the field. These characteristics can be seen in the first row of Fig.~\ref{fig:j_evol}, where we show the time evolution for case $j = 0$, considering $s = 0$ (left panel), $s = 1$ (central panel), and $s = 2$ (right panel), and different $\beta$ values. In all three cases, we notice a drastic change at longer times, with the signal oscillating at a frequency modulated by $\mu\sim\beta^{1/2}/2M$, a characteristic of massive fields in Schwarzschild. Moreover, notice that for a small $\beta$ the initial ringdown is still parametrized, as its characteristics are mostly due to the PSMs.

\subsection{Terms $j=1$ and oscillations in the tail}

For $j > 0$ as there is no longer any confinement of the perturbation due to the effective mass term the fields. As such, the perturbations should decay considerably faster than the $j=0$ case. In Schwarzschild, the standard picture is that we observe a power-law behavior at late times, i.e., $\propto t^{-(2\ell+\gamma)}$, where the factor $\gamma$ depending on initial conditions and spin $s$~\cite{gundlach1994late,PricePhysRevD.5.2419}.

In the second row of Fig.~\ref{fig:j_evol} we show the time evolution of the Gaussian packet for $j = 1$. In this case, we can see that after the ringdown stage, the usual tail in Schwarzschild is modified, showing an oscillatory behavior, still decaying (apparently) with $t^{-\gamma}$, at least for intermediate times. Such behavior is similar to what happens with massive fields in Schwarzschild-de Sitter spacetime~\cite{Correa:2024xki}, where the tail is replaced by the oscillatory behavior of the de Sitter models.

\subsection{Terms $j=2$ and deviations in the tail power law}

When analyzing the potential considering only $j = 2$ we can see that the corrections in the potential~\eqref{eq:potencial_simpli} are proportional to $r^{-2}$, similarly to the term proportional to $\ell(\ell+1)+\beta$. Since this term has an influence for the power law behavior of the tail, as we pointed previously, we should expect that the value of beta will cause some changes at this stage of the evolution. This can be seen directly in the third row of Fig.~\ref{fig:j_evol}, where we show the scalar, vector, and gravitational cases, with the extra term of $j = 2$ in the effective potential. We see that although the decay still follows a power law, which is slightly different from the Schwarzschild case for small values of $\beta$. For the scalar case (left panel of Fig.~\ref{fig:j_evol} in row 3) the effect is more pronounced. We note that the change will essentially depend on the comparison between the $\beta$ value and $\ell(\ell+1)$, due to the functional form of the potential modification.

\subsection{Terms $j=3,4...$ and the parametrized ringdown}

Finally, we have the terms with $j > 2$. In this case, the tail remains unchanged and the main effect is a parametric change in the ringdown. However, the particular case with $j = 3$ has a characteristic of being proportional to $r^{-3}$, which is the same order as the term proportional to the mass of the BH inside the brackets in the Eq.~\eqref{eq:potencial_simpli}. As a consequence of this, in some cases, we can expect that vector perturbations, which typically do not have this term, may behave similarly to gravitational or scalar perturbations (for the same value of $\ell$). However, the values of $\beta$ for this to occur are unrealistic, since we would need $\beta\sim {\cal O}(1)$, being inconsistent with a perturbative picture.

In the fourth row of Fig.~\ref{fig:j_evol} we show the Gaussian evolution for the case $j = 3$. As we pointed out, we can see that the main characteristic is a subtle change in the ringdown, whose magnitude essentially depends on how the correction value of the potential compares to the existing ones in Schwarzschild. For example, for a fixed value of $\beta$, we see that the scalar case (left panel of the fourth row in Fig.~\ref{fig:j_evol}) is the most impacted in this case.

  \begin{table*}
	\caption{QNMs extracted using the Prony method, compared to the ones obtained from the CF computation.}
	\centering
	\begin{tabular}{c | c | c || c | c } 
		\hline
		$\beta$ & $s = 1, \ell = 1, j = 3$ & $s = 1, \ell = 1, j = 3$ (CF) & $s = 2, \ell = 2, j = 3$ &  $s = 2, \ell = 2, j = 3$ (CF)\\  [0.5ex] 
		\hline\hline
		0.0  &   $0.247477-0.092382i$  & $0.248263 - 0.0924877 i$ & $0.374178-0.084071i$ & $0.373672 - 0.0889623 i$\\ 
		0.1  &   $0.252106-0.093004i$  & $0.252962 - 0.093097 i$ & $0.377515-0.089727i$ & $0.376593 - 0.0891505 i$\\
		0.2  &   $0.256634-0.092397i$  & $0.257607 - 0.0936848 i$ & $0.380382-0.089634i$ & $0.379508 - 0.0893427 i$\\
		0.3  &   $0.261437-0.094440i$  & $0.262199 - 0.0942514 i$  & $0.383440-0.090260i$ & $0.382415 - 0.0895386 i$ \\
		0.4  &   $0.266131-0.095119i$  & $0.266738 - 0.094797 i$  & $0.386401-0.090320i$ & $0.385316 - 0.0897376 i$ \\
		\hline
		\hline
	\end{tabular}
	\label{table:2}
\end{table*}

Finally, to illustrate how the QNMs vary parametrically, in TABLE~\ref{table:2} we show the values of QNMs extracted from time evolutions for each value of parameter $\beta$. We used the Prony method and compare it to the CF computation. We see that in both two cases, there is a slight increase in both the oscillation frequency, both imaginary and real parts, illustrating a smooth variation of the modes with the $\beta$ value.

\subsection{Negative beta and instability values}\label{subsec:instability}

\begin{figure*}
    \centering
    \includegraphics[width=0.33\textwidth]{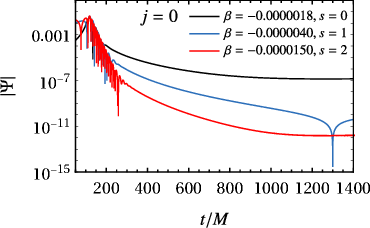}\includegraphics[width=0.33\textwidth]{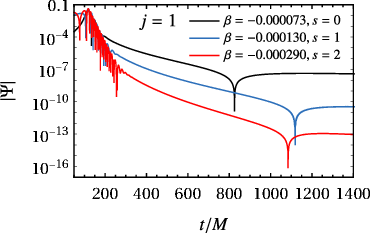}\includegraphics[width=0.33\textwidth]{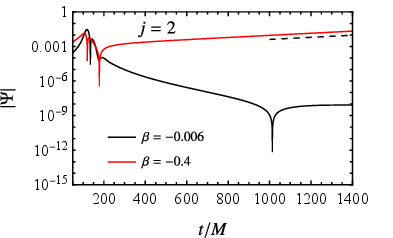}
    \caption{Time evolutions for negative values of $\beta_j$. For $j = 0$ (left panel), $\beta_j \sim -10^{-7}$ induces changes in the behavior of the time evolution. This also occurs, on a different scale, for $j = 1$ (central panel). The growth rate of the timescale depends on $\beta_j$, and in the case where $\ell = s = 0$, it already shows signs of growth for $\beta = -0.006$ (right panel). We also show the case $\beta=-0.4$, comparing it with the mode computed through the CF method (dashed line). }
    \label{fig:instability}
\end{figure*}

Now we turn our attention to negative values of $\beta$. A sufficient (but not necessary) condition for the instability can be imported from quantum mechanics studies~\cite{1995AmJPh..63..256B}, given by:
\begin{equation}
    \int_{-\infty}^\infty V dr_*<0.
    \label{eq:inequality}
\end{equation}
We see that this condition is always satisfied if $\beta_j$ is negative for $j=0$ (which incidentally indicates a tachyonic instability) and $j=1$. Therefore, we can conclude that for these corrections a negative $\beta_j$ always induces instabilities in the system. For $j>1$,
by using the effective potential described in this paper we find
\begin{equation}
\beta_j<-\frac{1}{2}(j-1)\left[\ell(\ell+1)+1-s^2\right].
\label{eq:inequality2}
\end{equation}
Eq.~\eqref{eq:inequality2} shows that there is always a negative value of $\beta_j$ that satisfies the sufficient condition for instability. It is important to note that the values of $\beta_j$ are not necessarily small in magnitude, which could, in principle, invalidate the perturbative characteristic of the expansion in the potential. However, since this condition is merely \textit{sufficient}, instability can still occur in cases where the potential expansion remains valid.

To illustrate the instability in the time-domain, in Fig.~\ref{fig:instability} we show some evolutions for negative values of $\beta_j$. Once again, we separate each specific term in the effective potential. We note that for the case $j=0$ (left panel) very small values of $\beta_j\sim-10^{-7}$ already induce changes in the behavior for the time-evolution on around. This also holds for $j=1$, but on a different scale (middle panel). It is worth noting that the growth rate depends on the value of $\beta_i$. From the right panel of Fig.~\ref{fig:instability} we see that the case $\ell=s=0$ already shows signs of growing for $\beta=-0.006$, even though the critical $\beta$ in \eqref{eq:inequality2} gives $-0.5$. This shows that even though the condition \eqref{eq:inequality} is not satisfied, one can still have instabilities in the system. It is worth noting that these unstable modes are not necessarily due to an already existing mode becoming unstable, as it happens with the $\ell=0$ mode in some beyond GR theories~\cite{Macedo:2020tbm}. Therefore, they may not be captured by tracking the evolution of the GR modes for negative $\beta$'s and, as such, are not parametrically connected to them. In the right panel of Fig.~\ref{fig:instability} we also show a direct comparison between the time evolution signal and the result from matching the computation of the unstable mode through the CF method $\omega=0.0020988 i$ (dashed line), which shows that the late time evolution is controlled by the purely imaginary unstable mode. 
\section{Conclusion}\label{sec:conclusion}

Black hole perturbation theory is an important tool to extract the behavior of the after-merger of two compact objects, including the late-time regime. We have investigated the ringdown behavior theories beyond GR, by modifying the effective potential for spherically symmetric black holes. While the parametrized modes were investigated in the past, there are many features that are not captured from pure expansions of the frequency around its GR values, such as tail modifications and the existence of new modes. Our work gives additional steps in those directions.

In this paper, we analyzed different types of additions in the effective potential, especially in the late time behavior, by looking into the time evolution of initial data. The signals clearly present distinctive features which could not be capture by looking into the parametrized computations of the modes in the frequency domain.

We note here that simple potential deviation is one form of testing beyond general relativity terms. More general setups should include coupling between the perturbations, which would induce mixing between the modes involved. Such features were already seem in some alternative theories of gravity, such as dynamical Chern-Simons~\cite{Molina:2010fb,Macedo:2018txb}. An interesting perspective would be to classify different types of coupling and their dynamical effects in gravitational waves signals.

After this paper was submitted for publication, Ref.~\cite{Thomopoulos:2025nuf} appeared on arXiv. Both studies are based on a master equation parametrization, incorporating a power-law correction to the effective potential. Additionally, while our approach includes spin fields s = 0, 1, 2, Ref.~\cite{Thomopoulos:2025nuf} focuses exclusively on the s = 2 sector, presenting a detailed analysis of the ringdown.

\begin{acknowledgements}
The authors thank Conselho Nacional de Desenvolvimento Científico e Tecnológico (CNPq), Coordenação de Aperfeiçoamento de Pessoal de Nível Superior (CAPES),  and Fundação Amazônia de Amparo a Estudos e Pesquisas (FAPESPA) for partial financial support.
\end{acknowledgements}

\bibliographystyle{spphys} 
\bibliography{svjourn3-epjc}

\end{document}